\documentclass[12pt]{article}
\usepackage{amsxtra,amssymb,amsthm,amsmath,latexsym}

\theoremstyle{plain}

\newcommand{\refS}[1]{Section~\ref{S:#1}}

\def\ds{\displaystyle}

\def\bysame{\rule{.5in}{.005in},\ }
\def\ve{\varepsilon}
\def\R{{\mathbb R}}

\def\calS{{\mathcal S}}
\def\calN{{\mathcal N}}
\def\calU{{\mathcal U}}
\def\calV{{\mathcal V}}
\def\jb{{\mathbf j}}

\def\kb{{\mathbf k}}
\def\l{\ell}
\def\bee{\begin{equation*}}
\def\eee{\end{equation*}}
\def\be{\begin{equation}}
\def\ee{\end{equation}}

\begin{document}
\title{Electromagnetic wave scattering by many small particles}

\author{A.G. Ramm\\
 Mathematics Department, Kansas State University, \\
 Manhattan, KS 66506-2602, USA\\
ramm@math.ksu.edu }

\date{}
\maketitle\thispagestyle{empty}

\begin{abstract} 


Scattering of electromagnetic waves by many small particles of arbitrary
shapes is reduced rigorously to solving linear algebraic system of
equations bypassing the usual usage of integral equations.

The matrix elements of this linear algebraic system have physical meaning. They
are expressed in terms of the electric and magnetic polarizability tensors.
Analytical formulas are given for calculation of these tensors with any desired
accuracy for homogeneous bodies of arbitrary shapes. An idea to create a 
"smart" material by embedding many small particles in a given region is
formulated.

\end{abstract}

\section{Introduction}\label{S:1}

Wave scattering by small particles was studied by Rayleigh, starting  
in 1871.  He understood that the main term in the field, scattered by a small
particle, is the dipole radiation.  A particle is small if $\lambda>>a$,
where $a$ is the characteristic dimension of the particle. The particle is
assumed homogeneous with parameters $\ve$, $\mu$ and $\sigma$. By
$k=\frac{2\pi}{\lambda}$ we denote the wave number in the medium,
surrounding particles.  There is a large literature on scattering by 
bodies, small in comparison with the wavelength (see \cite{D}, \cite{H},
\cite{L}, \cite{R476} and references therein). Exact analytical solutions
were found for spherical and ellipsoidal particles (\cite{L}).

For particles of arbitrary shapes the author gave analytical formulas for
S-matrix, which allow one to calculate this matrix with any desired 
accuracy (\cite{R476}). Our aim in this paper is
to show that the many-body scattering problem for $N$ small particles in
electromagnetic (EM) wave theory context can be rigorously reduced to
solving linear algebraic systems ({\bf las}) of equations bypassing the
usual usage of integral equations.

The matrix elements of this {\bf las} have physical meaning: they are
expressed in terms of the polarizability tensors of small bodies. The
decisive point is: the author has derived analytical formulas which allow
one to calculate these tensors with any desired accuracy for bodies of
arbitrary shapes. Our theory is developed in \refS{2}. The basic
assumption is: \be\label{e1} a<<\lambda<<d, \ee where $a$ is the
characteristic dimension of the small particles and $d$ is the smallest
distance between two distinct particles. In \refS{2} the problem is
formulated and a method of its solution is developed. 
Our theory in \refS{2} uses some ideas, similar to the ideas in \cite{F}.
The principal difference between the results in \cite{F} and 
in \refS{2} is the following: the scattering coefficients
in \cite{F} are not known analytically, and should be calculated 
separately, while in our theory the analogs of these coefficients, the 
scattering matrices $\calS_i$, are given analytically, explicitly
(see formulas (A.1)(A.5)-(A.7) in the Appendix). In \cite{F} the 
isotropic point scatterers are considered, the scattering is assumed 
isotropic, and in \refS{2} this assumption is
not used. However, this difference is less important than the 
principal difference, mentioned above.

In \refS{3} the
scattering in a medium consisting of many small particles is discussed
under assumption \eqref{e1}.  At the end of this Section we formulate an 
idea of creating a "smart" material nanotechnologically, by embedding many 
small particles in a given region in such a way that the resulting 
material would, for instance, have a desired radiation pattern.

In \refS{4} the field near the boundary of
small particles is discussed. In Appendix some auxiliary results from
\cite{R476} are given. These results are essential for our theory.

\section{Statement of the problem}\label{S:2}
Consider the scattering of a monochromatic plane EM wave with frequency 
$\omega$
by $N$ small homogeneous particles $D_i$, $1\leq i\leq N$, with parameters 
$\ve, \mu, \sigma$ (permittivity, magnetic permeability, conductivity) 
and Lipschitz boundaries $S_i$, placed in a medium with parameters 
$\ve_0, \mu_0, \sigma_0=0$. Let $S:=U^N_{i=1}S_i$, $D:=U^N_{i=1} D_i$. 
The time-dependent factor $e^{-i\omega t}$ is omitted.

For simplicity let us assume that $\sigma$ is so large that condition 
$[\calN,E]=0$ on $S$ holds, where $\calN$ is the exterior unit normal 
to $S$ and $[\calN,E]:=\calN\times E$ is the cross product.

Our theory can be developed for impedance boundary condition 
$[\calN,E]=\zeta [[~\calN,H],\calN]$ as well, where $\zeta$  is the 
surface 
impedance, $\zeta=\left(\frac{\ve'}{\mu}\right)^{1/2}$, and
$\ve':=\ve+\frac{i\sigma}{\omega}$. In this case the penetration depth 
$\delta$ of the EM field into the particle is given by the formula: 
$\delta=\sqrt{\frac{2}{\omega\sigma\mu}}$, and $\delta<< a$ if 
$\omega\sigma$ is sufficiently large, $\mu$ being fixed, 
$E\sim\frac{\delta}{\lambda}H$ if $\frac{\sigma}{\omega}>>\ve$.

The governing equations are
\be\label{e2}
  \nabla\times E=i\omega\mu H,
  \qquad \nabla\times H=-i\omega\ve'E\qquad\hbox{\ in\ }D, \ee
\be\label{e3}
  \nabla\times E=i\omega\mu_0 H,
  \qquad \nabla\times H=-i\omega\ve_0E\qquad\hbox{\ in\ }D', \ee
\be\label{e4}
  \hbox{$[\calN,E]$ and $\ve' E\cdot\calN$ are continuous across $S$,} \ee
\be\label{e5}
[\calN,E']=-[\calN,E_0]\qquad\hbox{\ on\ }S, \quad E=E_0+E', \quad 
H=H_0+H',\ee
where $E_0, H_0$ is the incident field, which satisfies \eqref{e3} in the 
whole space, $E', H'$ is the scattered field. Let
\be\label{e6}
  \calU:=\begin{pmatrix}E \\ H\end{pmatrix}, \ee
and $\calS_i$ be the scattering matrix corresponding to a small particle 
$D_i$. This means that if an electromagnetic wave $\calU$ is incident upon 
$D_i$, then the scattered field $\calU'$ in the far-field zone is
\be\label{e7}
  \calU'=\frac{e^{ikr}}{kr}\calS_i \calU+o\left(\frac{1}{r}\right),
  \qquad r=|x-x_i|>>\lambda>>a, \ee
where $x_i$ is a point in $D_i$, which can be chosen arbitrarily, and it 
does not matter which point is chosen since 
$D_i$ is small. Formula \eqref{e7} is valid if the field $\calU$ is 
practically homogeneous within the distances of order $a$. The foundation 
of our theory is an explicit formula for $\calS^{(n)}_i$ for a small 
homogeneous body of arbitrary shape, which allows one to calculate 
$\calS_i$ with any desired accuracy:
\be\label{e8}
  |\calS^{(n)}_i-\calS_i|=O(q^n), \qquad 0<q<1, \ee
where $q$ is a constant which depends only on the geometry of $\calS_i$ 
and the material 
constants $\ve_i,\mu_i,\sigma_i$. This formula is derived by the author 
(\cite[Chapter 7]{R476}) and the results we use in this paper are 
presented below, 
in the Appendix.

If $\calS_i$, $1\leq i\leq N$ are known and assumption \eqref{e1} holds, 
then the EM field at any point $x\in\R^3$, such that 
\be\label{e9}
  \min_{1\leq i\leq N} |x-x_i|\geq d, \ee
can be calculated by the formula:
\be\label{e10}
  \calU(x)=\calU_0(x)+\sum^N_{i=1} g(x,x_i) \calS_i\calV(x_i),
  \qquad g(x,y)=\frac{e^{ik|x-y|}}{k|x-y|}. \ee
The vectors $\calV(x_i)$  in \eqref{e10} are unknown. The expression
$ g(x,x_i)\calS_i\calV(x_i)$ is the field, scattered by $i$-th particle,
placed in the field $\calV(x_i)$.
If $\calV(x_i)$ were 
known, then formula \eqref{e10} would give the solution to the 
$N\hbox{-body}$ EM 
wave scattering problem in the region \eqref{e9}.
 
If $x$ is near the boundary $\calS_j$, then one gets:
\be\label{e11}
 \calU(x)=\calU_0(x)+\sum_{i\not= j}g(x,x_i)
 \calS_i\calV(x_i)+
 \begin{pmatrix}
   \nabla\times \ds\int_{\calS_j} g(x,s)\jb(s)ds\\
   \frac{1}{i\omega\mu}\nabla\times\nabla\times\ds\int_{S_j}g(x,s)\jb(s)ds
 \end{pmatrix}
\ee
In \eqref{e11} the integrals
represent the electric and magnetic fields generated by the $j\hbox{-th}$ 
particle in an immediate
neighborhood of this body, and $\jb$ is an unknown tangential field on
$\calS_j$, representing the induced by $\calU$ surface current.

If assumption \eqref{e1} holds, one may consider the $j\hbox{-th}$
particle as being placed in the homogeneous incident field
\be\label{e12}
\calU_0(x_j)+\sum_{i\not= j}g(x_j,x_i)\calS_i\calV(x_i),\ee
and the corresponding scattered field is
\be\label{e13}
\frac{e^{ik|x-x_j|}}{k|x-x_j|} \calS_j \left(\calU_0(x_j)
  +\sum_{i\not= j}g(x_j,x_i)\calS_i\calV(x_i)\right).\ee
Therefore we get \textit{ a linear algebraic system of equations} for the 
unknown
$\calV(x_i)$ in \eqref{e10}:
\be\label{e14}
  \calV(x_j)=
  \calU_0(x_j)+\sum_{i\not= j} g(x_j,x_i)\calS_i\calV(x_i),
  \quad 1\leq j\leq N. 
  \ee
If $\calS_j$, $1\leq j\leq N$, are known, then \eqref{e14} is a linear
system of 6N equations for the 6N unknowns $\calV(x_j)$, since
$\calV(x_j)$ is a 6-component vector. System \eqref{e14} can be solved
efficiently by iterations, provided that its matrix is diagonally
dominant. The matrix of system \eqref{e14} is

\be\label{e15}
  I-\sum_{i\not=j} \calS_i g(x_j,x_i), \ee where $I$ is the unit
matrix in the 6-dimensional space of vectors with complex-valued
coordinates and the norm of a vector is defined as follows:
\be\label{e16}
  \|\calU(x)\|:=\max_{1\leq i\leq6}|\calU_i(x)|,
  \qquad \calU=\begin{pmatrix}\calU_1\\ \vdots\\ \calU_6\end{pmatrix},
  \qquad E=\begin{pmatrix}\calU_1\\ \calU_2\\ \calU_3\end{pmatrix},
  \qquad H=\begin{pmatrix}\calU_4\\ \calU_5\\ \calU_6\end{pmatrix}.
   \ee
Matrix \eqref{e15} is diagonally dominant if
\be\label{e17}
  \|\sum_{i\not= j} \calS_i g(x_j,x_i)\|<1. \ee
Inequality \eqref{e17} holds if
\be\label{e18}
\frac 1 {kd}  \max_{1\leq m\leq 6} \sum^6_{\l=1}\sum_{i\not= j} 
  |(\calS_i)_{m\l}|<1.
\ee 
Here we have used the estimate of the norm of a matrix $A$, corresponding
to the norm \eqref{e16}:
\bee \|A\|=\max_{1\leq m\leq 6} \sum^6_{\l=1} |A_{m\l}|. \eee
Using formula (A.1) (see Appendix) in \eqref{e18} one gets
$ |(\calS_i)_{m\l}g(x_j, 
x_i)|=O(\frac{k^3V}{kd})=O(\frac{k^3a^3}{kd})<1$  
if assumption
\eqref{e1} holds. For example, if $ka\leq 0.1$, $kd=20$, $N<100$, then the
left-hand side of \eqref{e18} is less than $\frac{1}{20}\, 10^{-3}\,  
  6\cdot(N-1)\leq  3\cdot 10^{-2}<1,$
so that condition \eqref{e18} holds, and the system \eqref{e14} can be
efficiently solved by iterations:
\be\label{e19}
  \calV^{(n+1)}_j =
   \calU_{0j}+\sum_{i\not= j} g_{ji}\calS_i \calV^{(n)}_i,
  \qquad  \calU_{0j}:=\calU_0(x_j),\,\, g_{ji}:=g(x_j,x_i), \ee
\be\label{e20}
    \calV^{(0)}_j = \calU_{0j},  
\qquad \calV^{(n)}_i:=\calV^{(n)}(x_i).\ee
This completes the discussion of the scattering EM waves by N small
particles. In this discussion the field on the surface of $i\hbox{-th}$
particle was not calculated.

This field is not needed for  calculating  the vectors $\calU(x_j)$ 
which 
define the total field by formula \eqref{e10} at all points except 
immediate neighborhoods of the particles. See Section 4 for the
discussion of the field in an immediate neighborhoods of the particles. 

\section{Scattering in a medium consisting of small particles}\label{S:3}

If the number $N$ of particles is very large, say $N\sim 10^{23}$, so that
one discusses the EM wave scattering in a medium consisting of small
particles, then assumption \eqref{e1} implies that the number of small
particles per unit volume of the space is $O\left(\frac{1}{d^3}\right)$
and their total volume per unit volume of the space is
$O\left(\frac{a^3}{d^3}\right)\to 0$ if $\frac{a}{d}\to 0$.
Therefore the sum in \eqref{e10} in this limit tends to zero if assumption 
\eqref{e1} holds. Assumption \eqref{e1} was essential to our arguments, 
so, if $\ve,\sigma$ and $\omega$ are fixed, we arrive at the conclusion 
that scattering by $N$ small particles, under the assumptions 
\eqref{e1} and in the limit $\lim_{N\to\infty}\frac{a}{d}=0$, is vanishing 
in this limit.
However, the physical situation is dramatically different if we assume 
that $\ve=\ve(\omega)$ and $\sigma$ may depend in a suitable way on  
$a$ and $d$.

For example, assume that the small particles are 
identical spheres of radius $a$, $\sigma=0$. It is known that the  
polarizability tensor of a ball equals to
$$\alpha_{ij}=3\ve_0\,\frac{\ve-\ve_0}{\ve+2\ve_0}\delta_{ij},$$
where $\delta_{ij}$ is the Kronecker symbol, i.e., the unit matrix. 
Recall that the polarizability tensor $\alpha_{ij}$ of a dielectric
body $D$ with the dielectric constant $\ve$, placed in a
constant electrical field $E$ in a homogeneous 
medium with  the dielectric constant $\ve_0$, is defined by the 
equation:
$$P_i=\alpha_{ij}V\ve_0 E_j,$$
where $V$ is the volume of $D$,  over the repeated indices summation is 
understood, $P$ is the induced dipole moment acquired by the body $D$
in the field $E$, and $P_i$ is the $i$-th component of vector $P$.

Let us assume that the following limit exists:
\be\label{e21}
  \begin{aligned}
  \lim_{b\to 0}&\lim_{\nu\to\infty}
  \ \frac{\sum_{D_j\in B(y,b)}\calS_j}{\hbox{Vol}\,B(y,b)}:=q(y,\beta),\\
  & \hbox{Vol}\,B(y,b)=\frac43\pi b^3,
  \end{aligned}\ee
where the limit in \eqref{e21} is taken as the number $\nu$ of particles 
in the
ball $B(y,b)$ of radius $b$ centered at a point $y$ in $D$ tends to
infinity in such a way that assumption \eqref{e1} holds and
$\frac{a}{d}\to 0$, and
 $\beta$ is the scattering direction, $\beta=\frac{x-y}{|x-y|}$ if 
the scattering wave is directed from point $y$ to point $x$.

Equation \eqref{e10} in the limit
\be\label{e22}
  N\to\infty,\quad \frac{a}{d}\to 0,  
  \qquad a<<\lambda<<d, \quad k=\frac{2\pi}{\lambda},\ee
yields an integral equation for $\calU(x)$:
\be\label{e23}
  \calU(x)=\calU_0(x)+\int_D\ \frac{e^{ik|x-y|}}{k|x-y|} q 
(y,x)\calU(y)dy. \ee

If the particles are identical spheres of radius 
$a$, we have 
$$\nu\sim \frac43 \pi b^3/d^3,\quad \hbox{\quad and \quad} 
\alpha_{ij}=3\ve_0\ \frac{\ve-\ve_0}{\ve+2\ve_0}\delta_{ij},$$ 
so the nonzero limit \eqref{e21} exists if the limit
\be\label{e24}
  \lim_{\frac{a}{d}\to 0} \ \frac{a^3}{d^3}\ 
  \frac{\ve-\ve_0}{\ve+2\ve_0}:=w\not= 0\ee
exists. This limit exists and $w=\frac {\ve-\ve_0}{\kappa}$, if
\be\label{e25}
  \ve=-2\ve_0+\kappa\frac{a^3}{d^3}, \qquad \kappa\not=0,\ee
where $\kappa$ is a constant.
There also exist materials with $\ve=\ve(\omega)<0$ and $\sigma\neq 0$ for 
which the limit
\eqref{e25} exists. For example,
if $\ve:=\ve(\omega)+i\frac{\sigma}{\omega}$, $\sigma\neq 0$, and
\be\label{e26}
  \ve(\omega)\to -2\ve_0,
  \quad \frac{i\sigma}{\omega}\to \kappa_1\frac{a^3}{d^3}
  \quad\hbox{\ as\ }\omega\to\omega_0,
  \quad \kappa_1\neq 0,\ee
where $\omega_0$ is some frequency and $\kappa_1$ is a constant, then the 
nonzero limit \eqref{e24} 
exists and, therefore, the nonzero limit \eqref{e21} also exists.

Physically, it is natural to assume that the self-consistent (effective) 
field in a medium, consisting of many $(N\sim 10^{23})$ small particles, 
should not change if any single particle is removed from the medium, in 
spite of the fact that the exact field can be changed very much in an 
immediate neighborhood of the removed particle (\cite{R450}). For 
instance, if the boundary condition on the surface $S_i$ of this particle 
was $[\calN,E]=0$ then, after the particle is removed, $[\calN,E]\not=0$ 
on 
$S_i$, so the relative change is infinite.

 If the limit \eqref{e21} 
exists, then equation \eqref{e23} is the equation for the self-consistent 
field in the medium consisting of many small particles.

Our result allows one to formulate an idea for creating 
nanotechnologically a "smart" material by embedding many small particles 
in a given bounded region. Such an idea was first proposed for acoustic 
wave 
scattering in \cite{R492}, see also \cite{R512}.
The idea can be briefly described as follows. If the limit \eqref{e21}
exists, then the integral equation  \eqref{e23} holds. 
If for $x$
in the far zone $q(y, \beta):=q(y, x)=q(y)$, so that $q$ does not depend 
on $x$,
then equation 
 \eqref{e23} is equivalent to the Schr\"odinger equation with
potential $q$. The "smart" material, which is characterized by a desired
scattering amplitude (radiation pattern), can be constructed by 
first solving the inverse problem of finding such a $q$ from the 
desired scattering amplitude, that is, by solving an inverse scattering 
problem, and then 
relating this $q$ to the density of distribution of the small particles,
embedded in our region. If $N(y)$ is the number of the small particles
per unit volume in a neighborhood of the point $y$, then $q(y)$ is
proportional to $N(y)$ (see \cite{R492}). Thus, if $q$ is found, then 
$N(y)$ is known,
and the small particles should be embedded with the density $N(y)$ in 
order
that the resulting "smart" material would have the desired radiation 
pattern. Such is the situation in acoustic wave scattering by small 
bodies.
 The details are discussed in this case  
in \cite{R492} and \cite{R512}.

In EM wave scattering the potential $q$, in general, may depend on both 
variables, $y$ and $x$. This leads to a new inverse problem of finding 
such a $q=q(y,x)$ from the scattering data, an open problem 
currently.  Therefore, it is of interest to find if there are physically 
reasonable conditions under which $q(y,x)=q(y)$, because the inverse 
scattering problem in this case has been solved by the author (see 
\cite{R470}, chapter 5).

\section{Equation for the field in an immediate neighborhood 
of a particle}\label{S:4}

Formula \eqref{e10} was derived for $x$ satisfying condition \eqref{e9}. 
If one is interested in the field near $S_j$, one has to find $\jb$ in 
equation \eqref{e11} and then use formula  \eqref{e11} to calculate the 
field
at a point $x$ near $S_j$.
 
The electrical field near $S_j$ 
is
\be\label{e27}
  E(x)=E_0(x)+\sum_{i\not= j}E_i(x)+\nabla \times \int_{S_j} 
g(x,s)\jb(s)ds,\ee
where the vector $E_i(x)$ is calculated as the first three components of 
the vector $$\calS_i\calU(x_i) g(x_j, x_i),$$ see formula  \eqref{e11}.
The boundary condition $[\calN,E]\big|_{S_j}=0$ and formula 
\eqref{e27} 
yield an equation for $\jb$:
\be\label{e28}
  A\jb+\jb =\,-\frac{1}{2\pi}
  \left[\calN,E_0(s)+\sum_{i\not= j}E_i(s)\right],
  \quad s\in S_j, \quad 1\leq j\leq N,\ee
where
\be\label{e29}
  A\jb:=\int_{S_j} \left[\calN(s), 
  \left[\nabla_s \frac{e^{ik|s-t|}}{2\pi k|s-t|}, \jb(t)\right]\right] 
dt,
 \quad s\in S_j.\ee
Formula \eqref{e28} is analogous to formula (1.27) in \cite{R476}, 
being its generalization to the potentials of single layer with the 
density $\jb$ which is a tangential to $S_j$ vector field.

Equation \eqref{e28} is a Fredholm-type integral equation if $\calS_j$ is
a a sufficiently smooth surface. It allows one to find $\jb$ uniquely if 
$k^2$ is not an eigenvalue of the homogeneous Maxwell's equation in $D$ 
with the boundary condition  
$$[\calN,E]\big|_{S}=0.$$
If $k^2$ is such an 
eigenvalue, then this integral equation can be modified so that the 
modified equation is uniquely solvable for $\jb$ (see \cite{R196}).  

If $\jb$ is
found, then formula \eqref{e11} allows one to calculate
$\calU=\begin{pmatrix}E\\H\end{pmatrix}$ in a neighborhood of $S_j$ and on
$S_j$.

\section*{Appendix}
In this Appendix we collect some of the results from \cite{R476} used in 
\refS{2}.

The formula for the $\calS_E\hbox{-matrix}$ for EM wave-scattering by a 
single small body is (\cite[p.114]{R476}):

\bee
\calS_E=\frac{k^3V}{4\pi}
\begin{pmatrix}
\mu_0\beta_{11}+\alpha_{22}\cos\theta-\alpha_{32}\sin\theta 
  & \alpha_{21}\cos\theta-\alpha_{31}\sin\theta-\mu_0\beta_{12}\\
\alpha_{12}-\mu_0\beta_{21} \cos\theta+\mu_0\beta_{31}\sin\theta
  & \alpha_{11}+\mu_0\beta_{22}\cos\theta-\mu_0\beta_{32}\sin\theta
\end{pmatrix}_. \eqno{\hbox{(A.1)}}\eee
In (A.1) $\calS_E$ differs by the factor $k$ from the 
$\calS\hbox{-matrix}$ in \cite[p.114]{R476} because we use the 
dimensionless function $\frac{e^{ikr}}{kr}$, while  in \cite{R476} the 
function 
$\frac{e^{ikr}}{r}$ is used. The notations are the same 
as in \cite{R476}:

$\theta$ is the angle of scattering, 
$\alpha_{ij}=\alpha_{ij}(\gamma_\ve)$ is the electric polarizability 
tensor, $\gamma_\ve:=\frac{\ve-\ve_0}{\ve+\ve_0}$, 
$\beta_{ij}=\alpha_{ij}(-1)$ is the magnetic polarizability tensor, 
$V$ is the volume of the small body, and 
$\calS_E$ is defined by the formula for the scattered electrical field 
$E'$ in 
the far zone:
\bee
  E'=\frac{e^{ikr}}{kr}\calS_E E, \qquad r:=|x|, \eqno{\hbox{(A.2)}}\eee
where $E$ is the incident field at the point where the small body is 
located. If $E'$ is known, then
\bee
  H'= i\omega\mu_0 \frac{e^{ikr}}{r} \,ik[x^0,E'],
  \qquad  x^0=\frac{x}{|x|}. \eqno{\hbox{(A.3)}} \eee
Thus, if one knows $\calS_E$, then one can calculate the matrix $\calS$ in 
\eqref{e7}. 

The matrix $\calS_E$ in (A.1) is calculated in the coordinate 
system in which the wave vector $\kb$, $|\kb|=k,$ of the incident wave and 
vector $\kb'$ of the 
scattered wave are lying in one plane, $\theta$ is the angle between $\kb$ 
and $\kb'$, the $x'\hbox{-axis}$ coincides with the $x\hbox{-axis}$,
$y'\hbox{-axis}$ makes an angle $\theta$ with $y\hbox{-axis}$ ,
$z'\hbox{-axis}$ is directed along $\kb'$, and the origin $0$ is the same 
for $x,y,z$ and $x', y',z'$ coordinates. The origin lies inside the small 
body (particle), and the plane $y0z=y'0z'$ is called the scattering plane. 
One 
has $E=(E_1,E_2,0)$ in $xyz$ coordinates, $E'=(E'_1,E'_2,0)$ in $x'y'z'$ 
coordinates, and
\bee
  \calS_E \begin{pmatrix}E_2\\E_1\end{pmatrix}
  =\begin{pmatrix}E'_2\\E'_1\end{pmatrix}.
   \eqno{\hbox{(A.4)}} \eee
Let us give formulas (see \cite[pp.54-55]{R476}) for calculating the 
polarizability tensors 
$\alpha_{ij}(\gamma)$, $$\gamma:=(\ve-\ve_0)(\ve+\ve_0):=\gamma_\ve,$$
of a homogeneous body $D$ with boundary $S$, volume $V$, and permittivity 
$\ve$, placed in the medium with permittivity $\ve_0$. We have
\bee
  |\alpha^{(n)}_{ij}(\gamma)-\alpha_{ij}(\gamma)|
  =O(q^n), \qquad 0<q<1, \eqno{\hbox{(A.5)}} \eee
where the $n$-th approximation $\alpha^{(n)}_{ij}(\gamma_\ve)$ to 
$\alpha_{ij}(\gamma_\ve)$ is given by the formulas:
\bee
  \alpha^{(n)}_{ij}=\frac{2}{V} \sum^n_{m=0}\frac{(-1)^m}{(2\pi)^m}
  \ \frac{\gamma^{n+2}-\gamma^{m+1}}{\gamma-1}\ b^{(m)}_{ij},
  \qquad n>0, \eqno{\hbox{(A.6)}} \eee
and
\bee\begin{aligned}
  b^{(m)}_{ij}
  &:=\int_S \int_S dsdt\,\calN_i(t) \calN_j(s)
  \int_S\underbrace{\cdots}_{m-1} \int_S\ \frac{1}{r_{st_{m-1}}}
  \psi(t_1,t)\psi(t_2,t_1) \dots \psi(t_{m-1},t_{m-2})
  dt_1\dots dt_{m-1}, \\
  & \qquad \psi(t,s):=\frac{\partial}{\partial\calN_t}\ \frac{1}{r_{st}},
   \qquad r_{st}:=|s-t|, \\
  &  \qquad \alpha^{(1)}_{ij}(\gamma_\ve)=\alpha(\gamma_\ve+\gamma^2_\ve) 
  \delta_{ij}-\frac{\gamma^2_\ve}{\pi V}\ b^{(1)}_{ij}.
  \end{aligned} \eee
By $\calN_j(s)$ we denote the $j$-th component of the exterior unit normal
to $S$ at the point $s\in S$.

The magnetic polarizability tensor is:
\bee
  \beta_{ij}:=\alpha_{ij}(-1)+\alpha_{ij}(\gamma_\mu),
  \qquad \gamma_\mu:=\frac{\mu-\mu_0}{\mu+\mu_0}. \eqno{\hbox{(A.7)}}\eee
If $\mu=\mu_0$ then $\alpha_{ij}(\gamma_\mu)=0$. 

The induced electrical 
dipole moment $P$ on the body $D$, placed in a homogeneous electrostatic 
field $E$, is
\bee
  P_i=\alpha_{ij}(\gamma_\ve)V\ve_0 E_j.\eqno{\hbox{(A.8)}}\eee
Here and below summation is understood over the repeated indices.

The induced magnetic moment is
\bee
  M_i=\beta_{ij}V\mu_0H_j. \eqno{\hbox{(A.9)}} \eee
The first term in (A.7) is absent if the penetration depth $\delta$ of the 
magnetic field into $D$ is much larger than the size $a$ of $D$, and is 
present if $\delta<<a$, for example, if the body $D$ is a very good 
conductor in which $H$ cannot penetrate.

Finally, let us explain the remark below (A.2) about the 
$\calS\hbox{-matrix}$ defined in \eqref{e7}.
Matrix $\calS_E$ in (A.1) has been obtained in \cite{R476} from 
formula (7.95) in \cite{R476}. This matrix allows one to calculate the 
scattered field $E'$ by formula (A.4). If $E'$ is known, then the 
scattered field $H'$ can be calculated from the Maxwell equation 
\eqref{e3}:
\bee
  H'=\frac{\nabla\times E'}{i\omega\mu_0}=\frac{e^{ikr}}{r}
  \ \sqrt{\frac{\ve_0}{\mu_0}}\ [x^0,E'], \eee
which is formula (A.3), because $k=\omega \sqrt{\ve_0\mu_0}$.
Thus, the knowledge of the matrix (A.1) allows one to calculate 
$\calS\hbox{-matrix}$ in \eqref{e7}.

Let us discuss the role of the assumption \eqref{e1}. We have
\bee
  g:=\frac{e^{ik|x-y|}}{k|x-y|} 
  = \frac{e^{ik|x| \sqrt{1-  \frac{2(x^0,y)}{|x|}  + \frac{|y|^2}{|x|^2} }}}
  {k|x| \sqrt{1-\frac{2(x^0,y)}{|x|}+\frac{|y|^2}{|x|^2} }  },
 \quad x^0=\frac{x}{|x|}.\eee 
If $|y|<<|x|$, then
\bee
  g=\frac{e^{ik|x|}}{k|x|}  e^{-ik(x^0,y)}
   \left( 1+O\left(\frac{|y|}{|x|}\right)
           +O\left(\frac{k|y|^2}{|x|}\right)\right),
  \qquad k|y|<<1, \quad |y|<<|x|. \eee
If 
$$|y|\leq a,\,\, |x|\geq d,\,\, d>>a, \,\, a<<\lambda,\,\, 
\lambda=\frac{2\pi}{k},$$ 
then 
$$O\left(\frac{|y|}{|x|}\right)+O\left(k|y|\frac{|y|}{|x|}\right)<<1.$$
Thus, if $|x|\geq d$, $|y|\leq a$, $a<<d$ and $a<<\lambda$, then 
$$g\approx\frac{e^{ik|x|}}{|x|} e^{-ik(x^0,y)}.$$
The assumption $d>>\lambda$ is not used in these estimates. Therefore, on 
the wavelength $\lambda$ there can be many small particles as long as  
each of these particles is in 
the far zone with respect to all other particles. This was used in 
\cite{R450} 
in a study of the acoustic wave scattering in a medium consisting of many 
$(N\sim 10^{23})$ small, acoustically soft, particles.

In EM scattering one has to use the relation 
\bee 
  \nabla_x g=\frac{e^{ikr}}{kr} \left(ik-\frac{1}{r}\right)
  \ \frac{x-y}{|x-y|}, \qquad r=|x-y|, \eee
and in the far zone one neglects the term $\frac{1}{r}$ 
compared 
with $ik$. This can be done if  $$k>>\frac{1}{d}, \quad r\geq d,$$
and these inequalities imply  
$$d>>\lambda.$$
 That is why we have imposed assumption \eqref{e1} when dealing 
with EM wave scattering.

\end{document}